\documentclass[10pt]{article}          
\usepackage{mathrsfs,amsfonts,epsfig}
minus.4em \belowdisplayshortskip=2mm plus.2em minus.4em

\makeatletter 
\@addtoreset{equation}{section}  \makeatother

\title{\centerline{\normalsize  \hfill hep-ph/0605153}
{\bf Quantum oscillator on complex projective space (Lobachewski space) in
  constant magnetic field and the issue of generic boundary conditions }}

\author{ {\bf Pulak Ranjan Giri\thanks{e-mail :pulakranjan.giri@saha.ac.in}}\\
\normalsize Saha Institute of Nuclear Physics, 1/AF Bidhan-Nagar,  Calcutta
700064, India}

\date{\today}

\begin{document}\maketitle
\begin{abstract} \noindent\small
We perform a 1-parameter family of  self-adjoint extensions characterized by
the parameter $\omega_0$. This allows us to  get generic boundary conditions
for the quantum oscillator on $N$ dimensional complex projective
space($\mathbb{C}P^N$) and on  its non-compact version i.e., Lobachewski
space($\mathcal L_N$) in presence of constant magnetic field. As a result, we
get a family of energy spectrums for the oscillator. In our formulation  the
already known result of this oscillator is also belong to the family. We  have
also obtained energy spectrum which preserve all the symmetry (full hidden
symmetry and rotational symmetry)  of the oscillator. The method of
self-adjoint extensions  have been discussed for conic oscillator in presence
of constant magnetic field also.

\medskip

\end{abstract}
\section{\small{\bf {Introduction}}} \label{in}
Quantum oscillator plays a fundamental role in theoretical physics due to its
exact solvability and over-complete symmetry. The study of oscillator became
more interesting when the Euclidian oscillator was generalized on curved space
with constant curvature by P. W. Higgs. This generalized oscillator which is
known as Higgs Oscillator \cite{higgs} for obvious reason possesses lots of
interesting features.   For review see Ref. \cite{barut}.  The Euclidian
oscillator has also  been generalized on K\"{a}hler space  and various
properties of the system has been discussed in Ref. \cite{bellu}. In
Ref. \cite{ste}  exact solution of the quantum oscillator in $N$ dimensional
complex projective space ($\mathbb{C}P^N$),  Lobachewski space($ \mathcal
L_N$)  and related to cones in presence of  constant magnetic field has been
discussed. The relevance of this system to the higher dimensional quantum Hall
effect makes it interesting. It has been shown that the inclusion of constant
magnetic field does not break  any existing hidden symmetry of the oscillator
and super-integrability and exact solvability.

But  the solution of Ref. \cite{ste}  has been presented in terms of a fixed
boundary condition. As a consequence full symmetry of the energy spectrum has
not been obtained.  It is however possible to get generic boundary conditions
for the oscillator  by making a self-adjoint extensions of the Hamiltonian of
the system \cite{ste}. Consideration of the generic boundary  conditions are
not merely mathematical. It can be shown that the generic boundary conditions
help us to get a complete set of spectrums of the Hamiltonian under
consideration. In Ref. \cite{giri} such issue has been considered for quantum
mechanical oscillator on K\"{a}hler conifold in $2-$dimensions and it has been shown that the
consideration of self-adjoint extensions can help us to get energy spectrum,
which is degenerate with respect to the orbital and azimuthal quantum number.

In our present work we are going to address this issue for the oscillator
defined on $N-$dimensional complex projective space ($\mathbb{C}P^N$) and 
on  its non-compact version i.e., Lobachewski
space($\mathcal L_N$) in presence of constant magnetic field. We will perform a one
parameter family of self-adjoint extensions \cite{reed} of the initial domain
of the radial Hamiltonian of the harmonic oscillator \cite{ste} by von Neumann
method \cite{reed}.  This will help us to construct   generic boundary
conditions. We will show that for a specific value of the self-adjoint
extension parameter $\omega_0$ we can recover the known result \cite{ste}  and
for other values of the extension parameter $\omega_0$ we will get other
energy  spectrums which were not known so far. We will also discuss about the
degeneracy of the energy spectrum with respect to different quantum numbers,
which has been  possible for considering a one parameter family of
self-adjoint extensions of the radial Hamiltonian of the oscillator \cite{ste}.

However, the importance of self-adjointness of an unitary operator is far
fundamental. As we know evolution of a quantum system is dictated by unitary
group and the generator of that group is the Hamiltonian itself. According to
Stone's theorem \cite{reed} generators of unitary group (in this case
Hamiltonian) should be self-adjoint. So, for a non self-adjoint operator we
should search for a self-adjoint extensions if possible. If the system has
many self-adjoint extensions then different self-adjoint extensions should
unveil different physics for the system.

The paper is organized as follows: In Sec.~\ref{os}, we discuss the quantum
oscillator on complex projective space($\mathbb{C}P^N$) and  Lobachewski
space($\mathcal L_N$) in background constant magnetic field. In Sec.~\ref{ra},
we perform the self-adjoint  extensions  of the radial Hamiltonian of the
oscillator discussed in previous section and we make some observations for
some particular values of the extension parameter $\omega_0$. Here we show
that it is possible to retain complete degeneracy in the energy spectrum (full
hidden symmetry and rotational symmetry).  In Sec.~\ref{co}, the  method of
self-adjoint extensions has been discussed for  conic oscillator in constant
magnetic field.  We conclude  in Sec.~\ref{con}.

\section{\small{\bf {Quantum oscillator on $\mathbb{C}P^N$ and $\mathcal L_N$  
with background constant magnetic field}}} \label{os}

The quantum  oscillator on  complex projective space($\mathbb{C}P^N$) and on
Lobachewski space($\mathcal L_N$)  with background constant magnetic field $B$
is defined by the symplectic structure $\Omega $ and the Hamiltonian
$\widehat{\mathcal H}$  respectively as;
\begin{eqnarray}
 \Omega= d\pi_a\wedge dz_a+ d{\bar \pi}_a\wedge d{\bar z}^a+ iBg_{a\bar
 b}dz^a\wedge d{\bar z}^b \label{symp}\\ \widehat{\mathcal H}=
 \frac{1}{2}g^{a\bar{b}}(\hat{\pi_a}\hat{\bar{\pi_b}}+
 \hat{\bar{\pi_b}}\hat{\pi_a})+{\omega}^2g^{\bar{a}b}K_{\bar{a}}K_b,
\label{ham}
\end{eqnarray}
where the metric is of the form
\begin{eqnarray}
 g^{\bar a b}= \frac{2}{r^2_0}(1+\epsilon z\bar{z})(\delta^{ab}+\epsilon z^a
 {\bar z}^{b}),
\end{eqnarray}
and the  K\"{a}hler potential $K$  and its derivatives  $K_a$, $K_{\bar a}$
are given by
\begin{eqnarray}
K=\frac{r^2_0}{2\epsilon} \log (1+\epsilon z\bar z), \quad \epsilon=\pm 1,~
K_a=\frac{\partial K}{\partial z^a}=\frac{r_0^2}{2}\frac{\bar{z}^a}{1+\epsilon
z\bar z},~ K_{\bar a}= \frac{\partial K}{\partial
\bar{z}^a}=\frac{r_0^2}{2}\frac{z^a}{1+\epsilon z\bar z}.
\label{kpot}
\end{eqnarray}
The representation of the momentum operators $\pi_a$ and $\bar{\pi}_a$
consistent with the symplectic structure (\ref{symp}) take the forms  
\begin{eqnarray}
\hat\pi_a= -i(\hbar\partial_a+ \frac{B}{2}K_a),~~~ \hat{\bar{\pi}}_a=
-i(\hbar\partial_{\bar a} - \frac{B}{2}K_{\bar a}).
\label{momentum}
\end{eqnarray}
In order to investigate the maximum possible energy spectrums for the
oscillator,  let us consider the spectral problem
\begin{eqnarray}
{\widehat{\cal H}}\Psi= E\Psi,~~~\hat {J}_0\Psi= s\Psi,~~~\hat{\bf J}^2\Psi=
j(j+N-1)\Psi. \label{eigen}
\end{eqnarray}
It is convenient   to transform to the $2N$ dimensional spherical coordinates
$(r,\phi_i)$, where $i=1,......,2N-1$, $r$ is a dimensionless radial
coordinate taking values in the interval $[0,\infty)$  for $\epsilon=+1$, and
in $[0, 1]$ for $\epsilon=-1$ and $\phi_i$'s are appropriate angular
coordinates. In this spherical coordinates the above energy eigenvalue
equation in Eq. (\ref{eigen}) can be separated into radial coordinate if we
consider the trial wavefunction of the form
\begin{eqnarray}
\Psi=\psi(r) D^j_{s}(\phi_i), \label{wavef}
\end{eqnarray}
where $D^j_s(\phi_i)$ is the eigenvalue of the operators $\hat{\bf J}^2$,
$\hat{\bf J}_0$. It can be expressed via $2N$ dimensional Wigner functions,
$D^j_{s}(\phi_i)= \sum_{m_i} c_{m_i}D^j_{{m_i},s}(\phi_i)$, where $j,m_i$
denote the total and azimuthal angular momentum quantum number respectively.
\begin{eqnarray}
&&\widehat{J}_0 D^j_s(\phi_i)=s D^j_s(\phi_i),\label{J0}\\ &&\widehat{\bf J}^2
D^j_s(\phi_i) =j(j+N-1)D^j_s(\phi_i),\quad \widehat{J}_3 D^j_{m_i,s}=m_i
D^j_{m_i,s},\\ && m,s=-j,-j+1,\ldots , j-1, j\;\;  j=0,1/2,1,\ldots \label{mj}
\end{eqnarray}
%
%
Separating the differential equation we get the radial eigenvalue equation  of
the form
\begin{eqnarray}
H(r)\psi(r)= E\psi(r), \label{radialeigen}
\end{eqnarray}
where the radial Hamiltonian in Eq.  (\ref{radialeigen}) can be written in
spherical
coordinates as follows: 
\begin{eqnarray}H(r)=  
-\frac{\hbar^2}{2r^2_0}(1+\epsilon r^2)\left[ \frac{d^2}{d r^2} +
\frac{2N-1+\epsilon r^2}{r(1+\epsilon r^2)}\frac{d}{d r}+
\frac{4j(j+N-1)}{\epsilon r^2(1+\epsilon r^2)}+\right.\nonumber\\ \left.
\frac{\epsilon}{(1+\epsilon r^2)}(2s + \frac{\mu_B}{\epsilon})^2 -
\frac{\omega^2 r_0^4 r^2}{\hbar^2(1+\epsilon r^2)^2} + \frac{\epsilon
\mu_B^2}{(1+\epsilon r^2)^2}\right]\label{radham}
\end{eqnarray}
where
\begin{eqnarray}
r= \sqrt{z\bar z},~~~ \mu_B= \frac{B r_0^2}{2\hbar}.
\end{eqnarray}
and we have replaced $\hat{\bf J}^2$ and $\hat J_0$ by its eigenvalue
$j(j+N-1)$ and $s$ respectively.

We now move to the next section to discuss the self-adjointness of the radial
Hamiltonian $H(r)$ of  Eq.(\ref{radham}).

\section{\small{\bf {Self-adjointness of the radial Hamiltonian }}} \label{ra}
The effective radial Hamiltonian $H(r)$ of Eq.(\ref{radham}) is formally
self-adjoint, but formal self-adjointness does not mean that it is
self-adjoint on a  given domain \cite{dunford}. This operator $H(r)$ belongs
to unbounded differential operator defined on a Hilbert space. As we have
mentioned in our introduction, we will now perform self-adjoint extensions of
the operator $H(r)$ by von Neumann's method \cite{reed}. But before that let
us  briefly review here the von Neumann method  for the shake of completeness.

Let us consider an unbounded differential operator $T$ defined over a Hilbert
space $\mathcal H$ and consider a domain  $D(T)\subset \mathcal H$ for the
operator $T$ such that it becomes symmetric on the domain  $D(T)\subset
\mathcal H$. Note that the operator $T$ is called symmetric or Hermitian if
$(T\phi,\chi)= (\phi, T\chi)~~ \forall \phi,\chi \in D(T)$, where (.~,~.) is
the inner product defined over the Hilbert space $\mathcal H$. Let
$D(T^\dagger)$ be the domain of the corresponding  adjoint operator
$T^\dagger$. The operator $T$ is self-adjoint iff $T= T^\dagger$ and $D(T)=
D(T^\dagger)$.

We now state the criteria of self-adjointness of  a symmetric operator $T$
according to von Neumann method. We need to find out the the deficiency
subspaces (it is actually a null space) $D^\pm \equiv \mbox{Ker}(i\mp
T\dagger)$ and the deficiency indices $n^\pm(T) \equiv \dim(D^\pm)$.
Depending upon $n^\pm$,  $T$ is classified as \cite{reed}:
\begin{list}{\arabic{enumi})}{\usecounter{enumi}}

\item $T$ is essentially  self-adjoint, if $n^+= n^- = 0$.

\item $T$ has  $n^2$-parameter(real) family of self-adjoint extensions, if $n^+ = n^-=
n \ne 0$.

\item $T$ has no self-adjoint extensions, if $n^+\ne n^-$. In this case $T$ is
called maximally symmetric.
\end{list}

We now return to the discussion of our effective radial differential operator
$H(r)$. This operator  is symmetric in the domain,
\begin{eqnarray}
D(H(r)) = \{\phi(r): \parbox[t]{9cm}{\mbox{$\phi(r) = \phi'(r) = 0 $},
  absolutely continuous, square integrable over its full range with  measure
  \mbox{$d\mu$} \}\,,}
\label{domain1}
\end{eqnarray}
where $d\mu= \frac{r^{2N-1}}{(1+\epsilon r^2)^{2N-1}}dr$,  $\phi'(r)$ is the derivative of $\phi(r)$ with respect to $r$.  The
domain of the adjoint operator $H^\dagger(r)$, whose differential expression
is same as $H(r)$ due to formal self-adjointness, is given by
\begin{eqnarray}
D^\dagger(H(r)) = \{\phi(r): \parbox[t]{9cm}{ absolutely continuous, square
    integrable  on the  half  line  with  measure \mbox{$d\mu$} \}\,,}
\label{domain2}
\end{eqnarray}
 $H(r)$ is obviously not self-adjoint \cite{reed}, because
\begin{eqnarray}
D(H(r))\ne D(H^\dagger(r)).
\label{nonselfad}
\end{eqnarray}
So we may ask whether there is any possible self-adjoint extensions \cite{reed}
for the problem? To answer this question we need to investigate whether there
is any square-integrable solutions for the differential equations
\begin{eqnarray}
H(r)^\dagger \phi^\pm = \pm i\phi^\pm.
\label{imaginarysol}
\end{eqnarray}
The square-integrable solutions of Eq. (\ref{imaginarysol}), apart from
normalization are given by
\begin{eqnarray}
\phi^\pm=\cases{ D t^{\frac{c -2}{2}}(1-t)^{\frac{b^\pm + a^\pm -c}{2}}\;
_2F_1(a^\pm, b^\pm; c; t),\; & for $\epsilon=1$; \cr D t^{\frac{c
-2}{2}}(1-t)^{-\delta -2a^\pm - \frac{c}{2}+1}\; _2F_1(a^\pm, b^\pm, c; t) &
for $\epsilon= - 1$,}
\label{constant}\end{eqnarray}
where the constants $a^\pm= a(\pm i)$, $b^\pm= b(\pm i)$ and $c$ of the
Hypergeometric function \cite{abr} $_2F_1(a^\pm,b^\pm,c;t)$ are given in
general form as
\begin{eqnarray} 
\textstyle{a(k) = \frac{1}{2}\left(2j+N+ \epsilon\delta - \sqrt{ \frac{ 2
      r_0^2 k}{\epsilon\hbar^2} + N^2+ \frac{\omega^2 r_0^4}{\hbar^2} +
      \mu_B^2}\right), b(k) =\cases{- a(k) +\delta+j_1+1, &  for $\epsilon=1$;
      \cr a(k) +\delta,  & for $\epsilon= - 1$;}}
\label{cond1}
\end{eqnarray}
\begin{eqnarray}
c= j_1+1,  j_1 = 2j+N-1, \delta^2=  \frac{\omega^2 r_0^4}{\hbar^2} +
(2s+\frac{\mu_B}{\epsilon})^2,t= \cases{\frac{r^2}{1+r^2},&  for $\epsilon=1$;
\cr r^2, & for $\epsilon= - 1$;}
\label{cond2}
\end{eqnarray}
The existence of these complex eigenvalues  of $H(r)^\dagger$  signifies that
$H(r)$ is not self-adjoint. The solutions $\phi^\pm$ belong to the null space
$D^\pm$ of $H(r)^\dagger \mp i$, where $D^\pm \in D^\dagger(H)$. The dimensions
 $n^\pm$ of $D^\pm$ are known as deficiency indices  and are given by
\begin{eqnarray}
n^\pm =  \dim(D^\pm).
\label{deficiencyindices}
\end{eqnarray}
Since in our case the deficiency indices are $n^+ = n^- = 1$, we can get a
1-parameter family of self-adjoint extensions of $H(r)$. The self-adjoint
extensions of $H(r)$ are given by $H(r)^{\omega_0}$ with domain
$D(H(r)^{\omega_0})$, where
\begin{eqnarray}
D(H(r)^{\omega_0})= \{ \psi(r)= \phi(r)+ \phi^+(r) + e^{i\omega_0}\phi^-(r) :
    \phi(r)\in D(H(r)), \omega_0\in \mathbb{R} (\bmod 2\pi)\}.
\label{selfdomain}
\end{eqnarray}
The bound state  solutions of $H(r)^\omega$ are of the form
\begin{eqnarray}
\psi(r) =\cases{ C t^{\frac{c -2}{2}}(1-t)^{\frac{b  + a  -c}{2}} \; _2F_1(a,
b; c; t), & for  $\epsilon=1$; \cr C t^{\frac{c -2}{2}}(1-t)^{-\delta - 2a -
\frac{c}{2}+1} \; _2F_1(a, b, c; t), & for $\epsilon= - 1$;}
\label{beigenvalue}\end{eqnarray}
where $a= a(E)$, $b= b(E)$, $c$  and $t$ are given in general form in 
Eq. (\ref{cond1}) and
Eq. (\ref{cond2}).  $C$ is the normalization constant.  To find out the
eigenvalues we have to match the function $\psi(r)$ with the domain
Eq. (\ref{selfdomain}) at $r\rightarrow 0$. In the limit  $r\rightarrow 0$,
\begin{eqnarray}
\psi(r) \to \cases{C t^{\frac{c -2}{2}}(1-t)^{\frac{b  + a  -c}{2}}
\left[\Gamma_1(a,b,c) + (1-t)^{c-a-b}\Gamma_2(a,b,c) \right], & for
$\epsilon=1$; \cr C t^{\frac{c -2}{2}} \left[\Gamma_1(a,b,c) +
(1-t)^{1+\frac{c}{2}}\Gamma_2(a,b,c) \right], & for  $\epsilon= -1$;}
\label{matching1}
\end{eqnarray}
and
\begin{eqnarray}
\phi^+(r) + e^{i\omega_0}\phi^-(r) \to \cases{D t^{\frac{c
-2}{2}}(1-t)^{\frac{b + a -c}{2}} \left[\bar\Gamma_1 +
(1-t)^{c-a-b}\bar\Gamma_2 \right],  & for $\epsilon=1$; \cr D t^{\frac{c
-2}{2}} \left[\bar\Gamma_1 + (1-t)^{1+\frac{c}{2}}\bar\Gamma_2 \right], & for
$\epsilon= -1$;}
\label{matching2}
\end{eqnarray}

where for any three constants $m,n,p$,  $\Gamma(m,n,p)$'s are of the form
\begin{eqnarray}\textstyle{
\Gamma_1(m,n,p) = \frac{\Gamma(p) \Gamma(p-m-n) \Gamma(m+n-p+1) \Gamma(1-p)}
{\Gamma(p-m) \Gamma(p-n)\Gamma(n-p+1)\Gamma(m-p+1)},  \Gamma_2(m,n,p) =
\frac{\Gamma(p)\Gamma(m+n-p)\Gamma(p-m-n+1)\Gamma(1-p)}
{\Gamma(m)\Gamma(n)\Gamma(1-n)\Gamma(1-m)}}
\end{eqnarray}
and
\begin{eqnarray}
\bar\Gamma_1 = \Gamma_1(a^+,b^+,c) + e^{i\omega_0}\Gamma_1(a^-,b^-,c),~~~
\bar\Gamma_2 =  \Gamma_2(a^+,b^+,c) + e^{i\omega_0} \Gamma_2(a^-,b^-,c)
\end{eqnarray}
Now comparing the respective coefficients  in Eq. (\ref{matching1}) and
Eq. (\ref{matching2}) we get the eigenvalue equation,
\begin{eqnarray}
f(E)\equiv
    \frac{\Gamma(a)\Gamma(b)\Gamma(1-b)\Gamma(1-a)}{\Gamma(c-a)\Gamma(c-b)\Gamma(b-c+1)\Gamma(a-c+1)} = \mathcal
    M\frac{\cos(\beta +\omega_0/2)}{\cos(\alpha +\omega_0/2) },
\label{compare}
\end{eqnarray}
where
\begin{eqnarray}
\mathcal M = \frac{\left|\Gamma(c -a^\pm)\right|\left|\Gamma(c
-b^\pm)\right|\left|\Gamma(b^\pm -c +1)\right|\left|\Gamma(a^\pm -c
+1)\right|}{\left|\Gamma(a^\pm)\right|\left|\Gamma(b^\pm)\right|\left|\Gamma(1
-a^\pm)\right|\left|\Gamma(1 -b^\pm) \right|},
\end{eqnarray}
\begin{eqnarray}
\beta =\left|\arg\left(\Gamma(c -a^\pm)\right)\right|+\left|\arg\left(\Gamma(c
-b^\pm)\right)\right| + \left|\arg\left(\Gamma(b^\pm - c +1)\right)\right|
+\left|\arg\left(\Gamma(a^\pm - c +1)\right)\right| ,
\end{eqnarray}
\begin{eqnarray}
\alpha = \left|\arg\left(\Gamma(a^\pm)\right)\right| +\left|
  \arg\left(\Gamma(b^\pm)\right)\right| + \left|\arg\left(\Gamma(1
  -a^\pm)\right)\right| +\left| \arg\left(\Gamma(1 -b^\pm)\right)\right|\,.
\end{eqnarray}
The eigenvalue for general value of $\omega_0$ can be calculated
by plotting the graph of Eq. (\ref{compare}). We have plotted the graph of
Eq. (\ref{compare}) below for getting a complete understanding of the
behavior of the spectrum with respect to the self-adjoint extension parameter
$\omega_0$.  But we can immediately calculate the eigenvalue analytically at
least for some values of the extension parameter $\omega_0$ in the boundary
condition. So to appreciate constructing generalized boundary condition we now
investigate  some special cases.
\begin{figure}
\begin{center}
\includegraphics[width=0.6\textwidth, height=0.2\textheight]{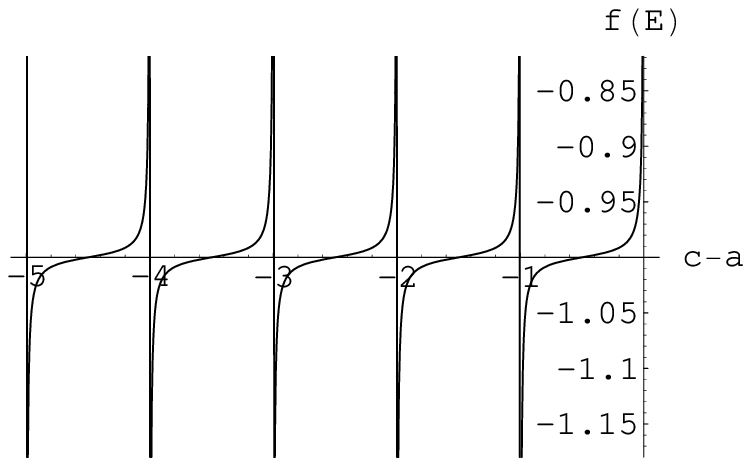}
\caption { \label{Figure2} A plot of Eq. (\ref{compare}) using
Mathematica with $N=2$, $\epsilon=-1$, $j=0$(actually we have taken the limit $j\to 0$, so
that the Eq. (\ref{compare}) make sense), $\delta= 0.001$ and energy range $c-a$ from $-5$ to $0$. The
horizontal axis  label by $c-a$ corresponds to the r.h.s =0 of Eq.
(\ref{compare}).}
\includegraphics[width=0.6\textwidth, height=0.2\textheight]{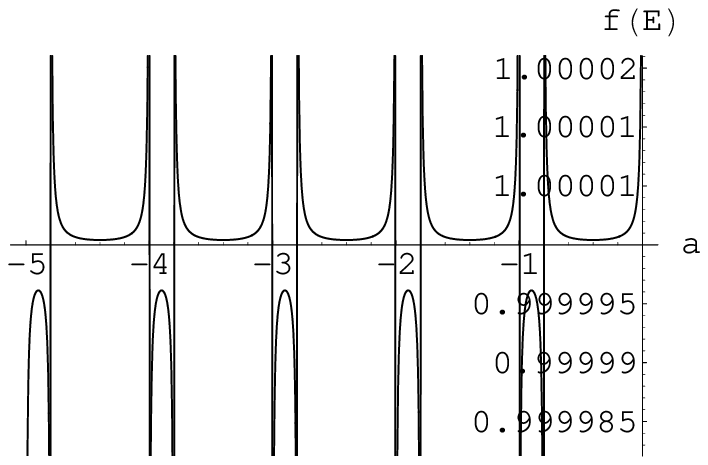}
\caption { \label{Figure1} A plot of Eq. (\ref{compare}) using
Mathematica with $N=2$, $\epsilon=+1$, $j=0$(actually we have taken the limit $j\to 0$, so
that the Eq. (\ref{compare}) make sense), $\delta= 1.2$ and energy range $a$ from $-5$ to $0$. The
horizontal axis  label by $a$ corresponds to the r.h.s $\neq 0$ of Eq.
(\ref{compare}).}
\end{center}
\end{figure}

\subsection{Case 1}
When the right hand side of Eq. (\ref{compare}) is infinity, we get $a = \pm
n$ or $b = \pm n$. $a= -n$ or $b= -n$ leads  to the eigenvalue, already
calculated in Ref. \cite{ste},
\begin{eqnarray}
E_{n,\,j,\,s} = \frac{\epsilon\hbar^2}{2r^2_0} \left[\left(2n+ 2j+ N +
    \epsilon \delta \right)^2  - (\frac{\omega^2 r_0^4}{\hbar^2} + N^2
    +\mu_B^2)\right],
\label{spectrum1}
\end{eqnarray}
%

%
The radial quantum number is given by
\begin{eqnarray}
n= \cases{0,1,\dots ,\infty& for $\epsilon=1$\cr  0,1,\dots ,n^{\rm
max}=[\delta/2-j-1]& for $\epsilon=-1$}
\label{quantumno}
\end{eqnarray}
For $a = +n$ and  $b= +n$ the energy spectrum will be the same expression
(\ref{spectrum1}), with $n$ replaced by $-n$.

%
%
\subsection{Case 2}
We can also make the right hand side of Eq. (\ref{compare}) zero, which gives
us $c-b= \pm n$ or $c-a =\pm n$. for $c-b= +n$, the energy spectrum becomes,
\begin{eqnarray}
E_{n,\,j,\,s} = \frac{\epsilon\hbar^2}{2r^2_0} \left[\left(2n - 2j - N +
\delta \right)^2  - (\frac{\omega^2 r_0^4}{\hbar^2} + N^2 + \mu_B^2)\right],
\label{spectrum4}
\end{eqnarray}
for $c-b=- n$, $n$ in (\ref{spectrum4}) will be replaced by $-n$ and radial
quantum number $n$ is given in (\ref{quantumno}). For $c-a =n$,
\begin{eqnarray}
E_{n,\,j,\,s} = \frac{\epsilon\hbar^2}{2r^2_0} \left[\left(2n- 2 j - N +
\epsilon\delta \right)^2  - (\frac{\omega^2 r_0^4}{\hbar^2} + N^2 +
\mu_B^2)\right],
\label{spectrum5}
\end{eqnarray}
For $c-a = -n$, $n$ in (\ref{spectrum5}) will be replaced by $-n$ and radial
quantum number $n$ is given in (\ref{quantumno}).

\subsection{Case 3}
For $c-b= +n+b$ and $c-a= +n+a$ we get
degenerate(degenerate with respect to orbital quantum no $j$) eigenvalue,
\begin{eqnarray}
E_{n,\,s} =  \frac{\hbar^2}{2r^2_0} \left[\left(n+ \delta  \right)^2  -
(\frac{\omega^2 r_0^4}{\hbar^2} + N^2 +\mu_B^2)\right], \mbox{for}
~~\epsilon=1.
\label{spectrum6}
\end{eqnarray}
For $c-b=-n +b$ and $c-a= -n+a$ we get,
\begin{eqnarray}
E_{n,\,s} = -\frac{\hbar^2}{2r^2_0} \left[\left(n + \delta  \right)^2  -
(\frac{\omega^2 r_0^4}{\hbar^2} + N^2 + \mu_B^2)\right], \mbox{for}
~~\epsilon= -1.
\label{spectrum7}
\end{eqnarray}
%
\subsection{Case 4}

Even if, we can get totally degenerate eigenvalue when  $c-b = c-a \pm n$  and
the form  of the spectrum is given by

\begin{eqnarray}
E_{n } =\frac{\hbar^2}{2r^2_0}\left[ n^2  - (\frac{\omega^2 r_0^4}{\hbar^2} +
N^2 + \mu_B^2)\right], \mbox{for}~~~\epsilon = +1\,.
\end{eqnarray}
For  $a+b+c=\pm n$ we get,
\begin{eqnarray}
E_{n } =-\frac{\hbar^2}{2r^2_0}\left[ n^2  - (\frac{\omega^2 r_0^4}{\hbar^2} +
N^2 + \mu_B^2)\right], \mbox{for}~~~\epsilon = -1\,.
\end{eqnarray}
\section{\small{\bf{Self-adjointness of conic oscillator in constant magnetic field}}}\label{co}
Study of self-adjointness of conic oscillator in constant background  magnetic
field is just a straightforward generalization of what we have done so
far. The $\nu -$ parametric family of cones over $\mathbb{C}P^N$ and $\mathcal
L_N$ is defined by the K\"{a}hler potential
\begin{eqnarray}
K=\frac{r^2_0}{2\epsilon} \log \left[1+\epsilon( z\bar z)^\nu\right],~~~\nu
>0;  \quad \epsilon=\pm 1,
\label{cokpotential} 
\end{eqnarray}
The metric is given by
\begin{eqnarray}
 g_{a\bar b}= \frac{\nu r_0^2 (z\bar z)^{\nu -1}}{2(1+\epsilon (z\bar z)^\nu
)} \left( \delta_{a b}-\frac{1-\nu+ \epsilon (z\bar z)^\nu}{z\bar z\;
(1+\epsilon (z\bar z)^\nu)}\bar z^a z^b\right),
\end{eqnarray}
The Hamiltonian of the system is same as Eq. (\ref {ham}). After doing some
algebra on energy eigenvalue equation of Eq. (\ref{eigen}),   we can arrive at
the radial Hamiltonian given by  Eq.(\ref{radham}) with
\begin{eqnarray}
\delta^2= \frac{\omega^2 r_0^4}{\hbar^2} + \left( 2\frac{s}{\nu} + \frac{B_0
r_0^2}{2\epsilon\hbar} \right)^2
\end{eqnarray}
and
\begin{eqnarray}
j_1^2 = \frac{(2j+N -1)^2}{\nu} + \frac{\nu-1}{\nu}\left[(N-1)^2 -
\frac{4s^2}{\nu}\right]
\end{eqnarray}
One can perform the self-adjoint extension of the radial Hamiltonian of  this
system also. Procedure is exactly the same as what we have done  above. 
Note that
the result will reduce to the result of Ref. \cite{giri} for magnetic field
$B= 0$ and $N=2$.
\section{\small{\bf {Discussion}}} \label{con}
The issue of self-adjointness, as pointed out in introduction, is of 
paramount importantace in quantum system due to Stone's theorem . 
It gurantees the spectrum to be the subset of real
line. Otherwise in principle the sectrum could be subset of the complex 
plane.  Complex eigenvalue could have importance in dissipative system.
However, in our work we have concentrated  on bound states of 
quantum oscillator on complex projective space($\mathbb{C}P^N$) and 
Lobachewski space($\mathcal L_N$) in background constant magnetic field. 
So Hamiltonian self-adjointness is must in our case.

We have obtained  a generic  boundary condition for the harmonic oscillator on
$\mathbb{C}P^N(\mathcal L_N)$ in constant magnetic field \cite{ste} and as a
result we have obtained a $\omega_0$-parameter family of energy eigenvalue
given by Eq. (\ref{compare}).There exists an energy spectrum at each point on
the circle $e^{i\omega_0}$.  We have shown that this generic boundary
condition can  restore the angular momentum degeneracy in energy spectrum for
a fixed value of the extension parameter $\omega_0$. In subsection $(3.3)$ we
have obtained eigenvalue which is independent of the orbital angular momentum
quantum number $j$. In subsection $(3.4)$ we have obtained eigenvalue which is
independent of both orbital and azimuthal quantum number. For consistency
checking,  we have also recovered the result of Ref. \cite{ste} in subsection
$(3.1)$. Not only that,  we have shown that it allows us to obtain more
solutions for different values of the extension parameter $\omega_0$, for
example we have calculated a case in subsection $(3.2)$. We have discussed the
conic oscillator in constant magnetic field background from the perspective of
self-adjointness of the system. It's a straight forward extension of what we
have done for oscillator on $\mathbb{C}P^N(\mathcal L_N)$ in constant magnetic
field.

\end{document}